\documentstyle[aps,tighten,preprint,epsfig,amssymb]{revtex}

\begin{document}

\draft

\preprint{nucl-th/0104068}


\title{
Asymmetries in $\bbox{\phi}$ photoproduction and the Okubo-Zweig-Iizuka
violation}


\author{Yongseok Oh\,$^a$%
\footnote{E-mail address : yoh@phya.yonsei.ac.kr}
and
H. C. Bhang\,$^b$}


\address{$^a$
Institute of Physics and Applied Physics,
Department of Physics, Yonsei University, Seoul 120-749, Korea \\
$^b$
Department of Physics, Seoul National University, Seoul 151-742, Korea}

\maketitle

\begin{abstract}

We study the vector meson density matrix and the associated
polarization asymmetries in $\phi$ photoproduction near threshold.
In order to study the OZI violating processes, we consider the direct
$\phi NN$ coupling as well as the knockout processes arising from the
non-vanishing $s\bar{s}$ sea quarks in the nucleon.
The polarization asymmetries in $\phi$ meson decays are found to be
very useful to constrain the $\phi NN$ coupling constant.
The hidden strangeness of the nucleon can also be studied by these
asymmetries, but mostly at large $|t|$ region.

\end{abstract}

\pacs{13.88.+e, 13.25.Gv, 13.60.Le, 24.70.+s}

\newpage

The electromagnetic production of $\phi$ meson from the nucleon has
been suggested as a sensitive probe to study the hidden strangeness of
the nucleon
\cite{HKPW91,HKW92,TOY94-97,TOY97,TOYM98,OTYM99-01}.
This is because the $\phi$ is nearly pure $s\bar s$ state so that
its direct coupling to the nucleon is suppressed by the OZI rule.
However, if there exists a non-vanishing $s\bar s$ sea quark component
in the nucleon, the strange sea quarks can contribute to the $\phi$
production via the OZI evasion processes.
Investigation of such processes can then be expected to shed light on
the nucleon strangeness content, if any \cite{EKKS00,Ellis01}.

In a series of publication, it has been argued that the double
polarization asymmetries of $\phi$ photo- and electroproduction from
the nucleon targets at threshold energy region can be used to probe
the hidden strangeness content of the nucleon
\cite{TOY97,TOYM98,OTYM99-01}.
In addition to the direct knockout production mechanisms, the direct
$\phi NN$ coupling can also induce an OZI evading process
\cite{Will98,TLT99,TLTS99}.
Even if we assume the ideal mixing of the $\phi$ with the $\omega$, the
effective $\phi NN$ coupling is allowed by the kaon loops and hyperon
excitations.
But the smallness of the effective $\phi NN$ coupling constants
estimated by this approach reveals the validity of the OZI rule
\cite{MMSV97}.
However the analyses of electromagnetic nucleon form factors
\cite{GHo76,HPSB76,Jaf89} and the baryon-baryon scattering
\cite{NRS75,NRS77,NRS79} suggest strongly a fairly large $\phi NN$
vector coupling constant which violates the OZI rule.
The recent CLAS experiments on $\phi$ photoproduction at large momentum
transfers \cite{CLAS00}, therefore, open another way to constrain
the $\phi NN$ coupling.
The data show increasing differential cross sections at large $|t|$,
which has been interpreted as the contributions from the $u$-channel
nucleon exchanges \cite{Lage00}.
It has been recently re-analysed to probe the nucleon resonances which
couple to the $\phi N$ channel in an SU(3) quark model
\cite{ZDGS99,ZSA01}.
The recent CLAS data on $\phi$ photo- and electroproduction
\cite{CLAS00,CLAS01} also lead to a discussion on the role of the axial
ghost pole in vector meson electromagnetic production at low energies
and large momentum transfers \cite{KV01}.

In this paper, we re-analyse the CLAS data by extending the work
of Refs. \cite{TOY97,TOYM98}.
As the production models, we consider the Pomeron exchange, $\pi$ and
$\eta$ exchange, nucleon pole terms with adjustable $\phi NN$ couplings,
and the direct knockouts.
We study the vector meson density matrix and the associated polarization
asymmetries in $\phi$ photoproduction and find that they are useful to
investigate the $\phi NN$ coupling as well as the knockout mechanisms.

\begin{figure}[t]
\centering
\epsfig{file=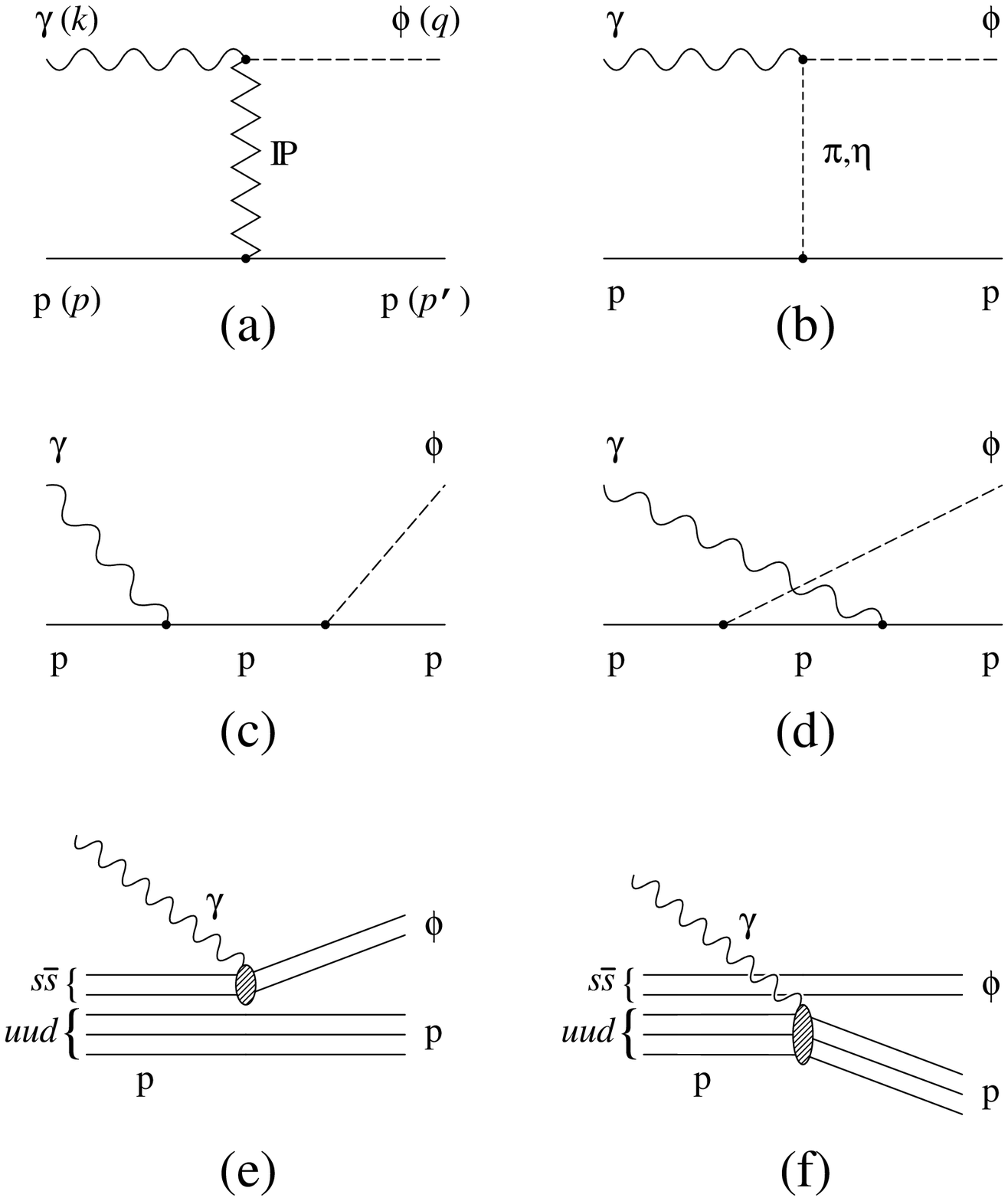, width=0.5\hsize}
\caption{
Diagrammatic representation of $\phi$ photoproduction mechanisms:
(a) Pomeron exchange, (b) ($\pi,\eta$) exchange, (c,d) nucleon pole
terms, and (e,f) direct knockout processes.}
\label{fig:diag}
\end{figure}

We start with defining the kinematic variables.
The four-momenta of the incoming photon, outgoing $\phi$ meson, and
initial/final proton are denoted by $k$, $q$, $p$, and $p'$,
respectively.
The Mandelstam variables are defined as $W^2 = (p+k)^2$ and
$t = (p-p')^2$.
The photon laboratory energy is denoted by $E_\gamma$.
We assume that the production amplitudes include the Pomeron exchange
[Fig.~\ref{fig:diag}(a)] and the pseudoscalar-meson $(\pi,\eta)$ exchange
[Fig.~\ref{fig:diag}(b)].
With the above $t$-channel background processes we study the direct
and crossed nucleon pole terms [Fig.~\ref{fig:diag}(c,d)] as well
as the direct knockout mechanisms [Fig.~\ref{fig:diag}(e,f)].

For the Pomeron exchange, which governs the total cross sections and
differential cross sections at low $|t|$,
we follow the Donnachie-Landshoff model \cite{DL84,DL86,DL87a,DL92},
which gives \cite{LM95,PL97}
\begin{equation}
T_{fi}^P = i M^{}_0 (s,t) \, \bar{u}^{}_{m_f^{}} (p')
\varepsilon^*_{\mu}(\phi)
\left\{ k \!\!\!/ g^{\mu\nu} - k^\mu \gamma^\nu \right\}
\varepsilon_{\nu}(\gamma) u^{}_{m_i^{}}(p),
\end{equation}
where $\varepsilon^{}_{\mu} (\phi)$ and $\varepsilon^{}_{\nu}
(\gamma)$
are the polarization vectors of the $\phi$ meson and photon,
respectively, and $u^{}_m(p)$ is the Dirac spinor of the nucleon with
momentum $p$ and spin projection $m$.
The Pomeron exchange is described by the following Regge
parameterization:
\begin{equation}
M^{}_0 (s,t)= C^{}_V \, F^{}_1(t) \, F^{}_V (t)\,
\left(\frac{s}{s_0} \right)^{\alpha_P^{} (t)-1}
\exp\left\{ - \frac{i\pi}{2} [ \alpha_P^{}(t) - 1] \right\},
\label{Pom:M0}
\end{equation}
where $F_1^{}(t)$ is the isoscalar electromagnetic form factor of the
nucleon and $F_V^{}(t)$ is the form factor for the
vector-meson--photon--Pomeron coupling:
\begin{equation}
F_1^{} (t) = \frac{ 4 M_N^2-2.8t }{ (4M_N^2-t) (1-t/t_0)^2 },
\qquad
F_V^{} (t) = \frac{1}{1 - t/M_V^2}\,
\frac{2\mu_0^2}{2 \mu_0^2 + M_V^2 - t},
\end{equation}
where $t_0 = 0.7$ GeV$^2$ and $M_N^{}$ ($M_V^{}$) is the nucleon
($\phi$ meson) mass.
The Pomeron trajectory is known to be $\alpha_P^{} (t) = 1.08 +
0.25\,t$.
The strength factor $C_V^{}$ reads $C_V^{} = 12 \sqrt{4\pi\alpha_{\rm
em}} \beta_0 \beta_s /f_V^{}$ with the vector meson decay constant
$f_V$ ($=13.13$ for the $\phi$ meson) and $\alpha^{}_{\rm em} = e^2/4\pi$.
By fitting all of the total cross section data for $\omega$, $\rho$, and
$\phi$ photoproduction at high energies, the remaining parameters of the
model are determined:
$\mu_0^2 = 1.1$ GeV$^2$, $\beta_0^{} = 2.05$ GeV$^{-1}$, $\beta_s^{} =
1.60$ GeV$^{-1}$, and $s_0 = 4$ GeV$^2$.

The pseudoscalar-meson exchange is known to be an important correction
to the Pomeron exchange at low energies. The amplitude can be calculated
from the following effective Lagrangians,
\begin{eqnarray}
{\mathcal{L}}_{\phi \gamma \Pi}^{} &=&
\frac{e g_{\phi\gamma \Pi}^{}}{M_V} \epsilon^{\mu\nu\alpha\beta}
\partial_\mu \phi_\nu \partial_\alpha A_\beta\, \Pi,\qquad
\nonumber\\
{\mathcal{L}}_{\Pi NN}^{} &=&
\frac{g_{\pi NN}}{2M_N}
\bar N \gamma^\mu \gamma_5\tau_3 N \partial_\mu \pi^0
+ \frac{g_{\eta NN}}{2M_N} \bar N \gamma^\mu \gamma_5 N \partial_\mu \eta,
\end{eqnarray}
where $\Pi = (\pi^0,\eta)$ and $A_\beta$ is the photon field.
The resulting invariant amplitude reads
\begin{eqnarray}
T^{ps}_{fi} = \sum_{\Pi=\pi,\eta}
\frac{-iF_{\Pi NN}(t) F_{\phi\gamma\Pi}(t)}{t-M_\Pi^2}
\frac{e g^{}_{\phi\gamma\Pi} g^{}_{\Pi NN}}{M_V} \,
\bar{u}_{m^{}_f}(p') \gamma_5 u_{m^{}_i}(p) \,
\varepsilon^{\mu\nu\alpha\beta} q_{\mu}^{} k^{}_\alpha
\varepsilon_\nu^* (\phi) \varepsilon^{}_\beta (\gamma).
\label{T:ps}
\end{eqnarray}
In the above, we have followed Ref. \cite{FS96} to include the 
form factors to dress the $\Pi NN$ and $\phi\gamma\Pi$ vertices as
\begin{equation}
F_{\Pi NN}^{} (t) = \frac{\Lambda_\Pi^2 - M^2_\Pi}
{\Lambda_\Pi^2 -t},  \qquad
F_{\phi\gamma\Pi}^{} (t) =
\frac{\Lambda_{\phi\gamma\Pi}^2-M_\Pi^2}
{\Lambda_{\phi\gamma\Pi}^2-t} .
\label{PS:FF}
\end{equation}
We use $g_{\pi NN}^2/4\pi = 14$ for the $\pi NN$ coupling constant.
The $\eta NN$ coupling constant is not well determined \cite{TBK94}.
In the case of $\omega$ photoproduction, the higher mass of
the $\eta$ and the smallness of the $\omega\gamma\eta$ coupling constant
suppress the contribution of the $\eta$ exchange compared with that of
the $\pi$ exchange \cite{OTL01}. 
However in the case of $\phi$ photoproduction, the $\eta$ exchange
contribution may be larger than the $\pi$ exchange at large $|t|$
depending on the $\eta NN$ coupling constant because of the large
value of the $\phi\gamma\eta$ coupling.
The dependence of the cross sections on $g_{\eta NN}^{}$ was discussed
in detail by Ref. \cite{TLTS99} and in this study
we use $g^2_{\eta NN}/4\pi = 0.99$ which is obtained from
making use of the SU(3) symmetry relation together with
a recent value of $F/D = 0.575$.
The coupling constants $g_{\phi\gamma\Pi}$ can be estimated
through the decay widths of $\phi \to \gamma\pi$ and
$\phi \to \gamma \eta$ \cite{PDG00}, which lead to
$g_{\phi\gamma\pi} = 0.14$ and $g_{\phi\gamma\eta} = 0.704$.
The amplitudes are also dependent of the cutoff parameters $\Lambda_\Pi$
and $\Lambda_{\phi\gamma\Pi}$ in Eq. (\ref{PS:FF}), which are fitted by
the experimental data \cite{CLAS00,BHKK74} at small and medium $|t|$
region and from Ref. \cite{TLTS99}:
$\Lambda_\pi^{} = 0.9$ GeV,
$\Lambda_{\phi\gamma\pi}^{} = 0.9$ GeV, $\Lambda_\eta^{} = 1.0$ GeV
and $\Lambda_{\phi\gamma\eta}^{} = 0.9$ GeV.

We evaluate the nucleon pole terms shown in Fig.~\ref{fig:diag}(c,d)
from the following interaction Lagrangians,
\begin{eqnarray}
{\mathcal{L}}_{\gamma NN}^{} & = &
- e \bar{N} \left( \gamma_\mu \frac{1+\tau_3}{2} {A}^\mu
- \frac{\kappa_N^{}}{2M_N^{}} \sigma^{\mu\nu} \partial_\nu A_\mu \right)
  N,
\nonumber\\
{\mathcal{L}}_{\phi NN}^{} & = &
- g_{\phi NN}^{} \bar{N} \left( \gamma_\mu {\phi}^\mu
- \frac{\kappa_\phi}{2M_N^{}} \sigma^{\mu\nu}
\partial_\nu \phi_\mu \right) N,
\end{eqnarray}
with the anomalous magnetic moment of the nucleon $\kappa_{p(n)} =
1.79$ $(-1.91)$.
There are big uncertainties with the $\phi NN$ coupling constants.
First, the effective $\phi NN$ coupling through kaon loops and hyperon
excitations was estimated to be very small, $g_{\phi NN} \simeq -0.24$
\cite{MMSV97}, which is consistent with the OZI rule prediction.
However the analyses on the electromagnetic nucleon form factors
\cite{GHo76,HPSB76,Jaf89} and the baryon-baryon scattering
\cite{NRS75,NRS77,NRS79} favor a large $\phi NN$
vector coupling constant which strongly violates the OZI rule:
$g_{\phi NN}^{} / g_{\omega NN} = -0.3 \sim -0.43$, therefore
$g_{\phi NN}^{} = -2.3 \sim -4.7$ with
$g_{\omega NN}^{} = 7.0 \sim 11.0$ \cite{SL96}.
In our study the nucleon pole terms are responsible to the recent CLAS
data on $\phi$ photoproduction at large $|t|$ and we fit the data with the
$g_{\phi NN}^{}$ and find that $|g_{\phi NN}^{}| = 3.0$ can reproduce
the data, which is consistent with the Regge trajectory study of Ref.
\cite{Lage00}.
The tensor coupling constant $\kappa_\phi$ has been estimated to be
$\kappa_\phi / \kappa_\omega = 0.5 \sim 2.1$
\cite{Jaf89,NRS75} or $\kappa_\phi \simeq \pm 0.2$ \cite{MMSV97}.
However, it is known that the value of the tensor coupling
$\kappa_\omega$ is very small as discussed in a study of $\pi N$
scattering and pion photoproduction \cite{SL96}.
Thus we use $\kappa_\phi = 0$ in our calculation.
We also found that the results are not dependent on $\kappa_\phi$
if $\kappa_\phi < 1$.
The resulting invariant amplitude reads
\begin{equation}
T^N_{fi} = \bar{u}_{m^{}_f} (p') \varepsilon^{\mu *} (\phi)
M_{\mu\nu} \varepsilon^\nu (\gamma) u_{m^{}_i}(p),
\label{T:N}
\end{equation}
where
\begin{eqnarray}
M_{\mu\nu} = -e g_{\phi NN}^{}
\left[ \Gamma_\mu^\phi(q)
\frac{\not\! p \ + \not\! k \ + M_N}{s-M_N^2}
\Gamma_\nu^\gamma(k) F_N^{} (s) +
\Gamma_\nu^\gamma(k)
\frac{\not\! p \ - \not\! q \ + M_N}{u-M_N^2}
\Gamma_\mu^\phi(q) F_N^{} (u)
\right]
\label{N-term}
\end{eqnarray}
with
\begin{equation}
\Gamma_\mu^\phi (q) = \gamma_\mu - i \frac{\kappa_\phi}{2M_N}
\sigma_{\mu\alpha} q^\alpha, \qquad
\Gamma_\nu^\gamma (k) = \gamma_\nu + i \frac{\kappa_p}{2M_N}
\sigma_{\nu\beta} k^\beta,
\label{Gamma-N}
\end{equation}
and $s= (p+k)^2$, $u = (p-q)^2$.
Here we follow Ref. \cite{HBMF98a} to include a form factor
\begin{eqnarray}
F_N (r) = \frac{\Lambda_N^4}{\Lambda_N^4  - (r - M_N^2)^2}
\label{N:FF}
\end{eqnarray}
with $r = s$ or $t$.
The gauge invariance is restored by projecting out the gauge
non-invariant parts \cite{OTL01} and we use
the cutoff parameter $\Lambda_N = 0.5$ GeV as in the study of $\omega$
photoproduction \cite{OTL01}.

If we assume a non-vanishing strange sea quarks in the nucleon
wavefunction, it can contribute to $\phi$ photoproduction through direct
knockout mechanisms \cite{HKW92} [Fig.~\ref{fig:diag}(e,f)].%
\footnote{The direct knockout mechanisms may be related to the loop
diagrams of the effective $\phi NN$ vertex considered in Ref.
\cite{MMSV97}, i.e., the meson clouds of the nucleon or the intermediate
baryons may interact with the photon to produce the $\phi$ meson. Since
we are considering tree diagrams only, however, clarifying that issue is
beyond the scope of this work.}
In order to investigate the effects from the hidden strangeness content
of the proton in $\phi$ photoproduction, we parameterize
the proton wavefunction in Fock space as
\begin{equation}
|\, p \rangle = A_0 |\, uud \rangle + \sum_X A_X |\, uud X \rangle
+ \sum_X B_X |\, uuds\bar s X \rangle,
\label{proton}
\end{equation}
where $X$ denotes any combination of gluons and light quark pairs of $u$
and $d$ quarks.
Ellis {\it et al.\/} \cite{EKKS95} estimated it to be
1--19\% from an analysis of $p \bar p$ annihilation.
{}From the $\phi$ electroproduction process, Henley {\it et al.\/}
\cite{HKW92} claimed that its theoretical upper-bound would be 10--20\%.
By employing a relativistic quark model \cite{TOY94-97} it was shown
that the upper-bound could be lowered to 3--5\%.
For simplicity and for our qualitative study, we approximate the proton
wavefunction (\ref{proton}) as
\begin{eqnarray}
|\, p \rangle &=& A |\, uud \rangle + B |\, uud s\bar s \rangle
\nonumber \\
&=& A | [uud]^{1/2} \rangle +
\sum_{j_{s\bar s} = 0,1; \, j_c} b_{j_{s\bar s}}
| \left[ \bbox{[} [uud]^{1/2} \otimes [{\bf L}]\bbox{]}^{j_c}
 \otimes [s\bar s]^{j_{s\bar s}} \right]^{1/2}
 \rangle ,
\end{eqnarray}
where ${\bf L}$ is the relative angular momentum of the two clusters.
This parameterization of the nucleon wavefunction can be justified in
$\phi$ production as argued in Refs. \cite{HKW92,TOY94-97}.
The strangeness content $B^2$ should be determined by further
theoretical and experimental studies and we present our results with
varying this value.
The production amplitude of this mechanism has been discussed
extensively in Refs. \cite{TOY94-97,TOYM98} and will not be repeated here.

\begin{figure}
\centering
\epsfig{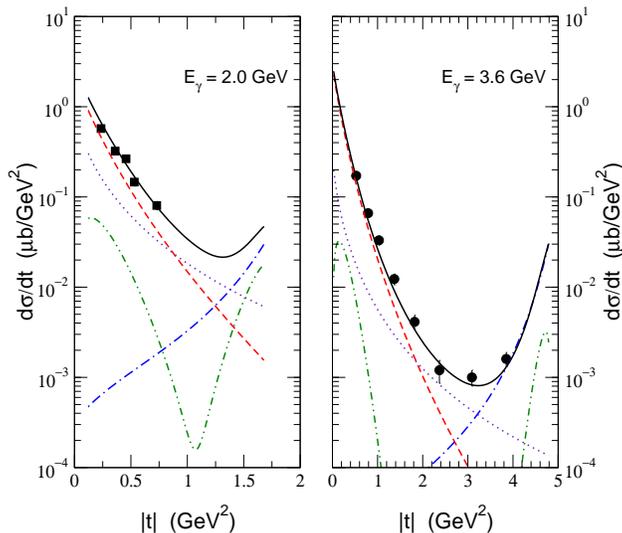}
\caption{Differential cross sections of $\phi$ photoproduction at
$E_\gamma = $ $2.0$ GeV and $3.6$ GeV.
The results are from Pomeron exchange (dashed), pseudoscalar-meson
exchange (dotted), nucleon pole terms with $g_{\phi NN}^{} =
-3.0$ (dot-dashed), direct knockouts with $B^2 = 1$ \% (dot-dot-dashed),
and the full amplitude (solid).
The experimental data are from Ref. \protect\cite{BHKK74} (filled squares)
and Ref. \protect\cite{CLAS00} (filled circles).}
\label{fig:dsdt}
\end{figure}

Within the models discussed above, we present the differential cross
sections of $\phi$ photoproduction at $E_\gamma = 2.0$ GeV and $3.6$ GeV
in Fig. \ref{fig:dsdt}.
In the considered energy region, it can be seen that the Pomeron
exchange (dashed lines) gives the dominant contribution at small $|t|$
region.
The pseudoscalar-meson exchange (dotted line) is important at
medium $|t|$ region while the nucleon pole terms are responsible
for the increase of the differential cross sections at large $|t|$.
With $|g_{\phi NN}^{}| = 3.0$, we could reproduce the recent CLAS data
on $\phi$ photoproduction \cite{CLAS00}.
This is consistent with the results of Ref. \cite{Lage00} based on
Regge trajectory.
The direct knockout mechanism is suppressed and thus cannot be
distinguished from the other mechanisms unless we study other quantities
such as the polarization asymmetries \cite{TOY97,TOYM98}.

\begin{figure}[t]
\centering
\epsfig{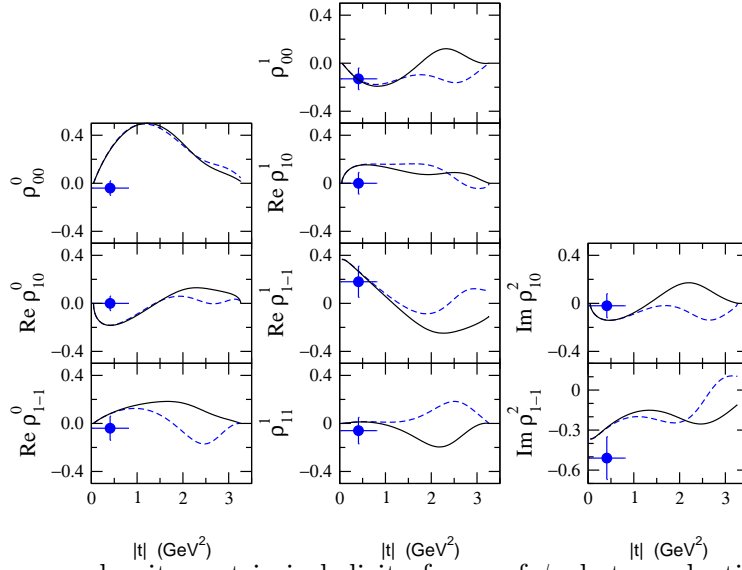}
\caption{Vector meson density matrix in helicity frame of
$\phi$ photoproduction at $E_\gamma = 2.8$ GeV with the full amplitude. 
The solid and dashed lines are obtained with $g_{\phi NN}^{} = -3.0$ and
$3.0$, respectively.
The data are from Ref. \protect\cite{LAMP2-78} at $E_\gamma = 2.8$ and $4.8$
GeV.}
\label{fig:dm2_8}
\end{figure}

Although the analyses on electromagnetic nucleon form factor and
baryon-baryon scattering supports the large value of the $\phi NN$
coupling, one can further constrain the coupling in $\phi$
photoproduction.
The electromagnetic nucleon form factor study based on vector meson
dominance only tells that the relative phase between the
$\phi \gamma$ coupling and the $\phi NN$ coupling is negative.
We found that the asymmetries in $\phi$ photoproduction can give us a
clue on the phase of the $\phi NN$ coupling and therefore the vector
meson dominance in electromagnetic nucleon form factor can be tested
directly.
(In our convention, the $\phi\gamma$ coupling has positive sign.)
For this purpose, we compute the $\phi$ meson density matrix elements
within our model.
The results at $E_\gamma = 2.8$ GeV in {\em helicity frame\/}
\cite{SSW70} are shown in Fig. \ref{fig:dm2_8}.
The solid lines are obtained with $g_{\phi NN}^{} = -3.0$ while the
dashed lines are with $g_{\phi NN}^{} = +3.0$.
The contributions from the direct knockout mechanisms are small and
cannot be seen here.
Figure \ref{fig:dm2_8} shows that the vector meson density matrix is
strongly dependent on the phase of the $g_{\phi NN}^{}$ at large $|t|$.

\begin{figure}[t]
\centering
\epsfig{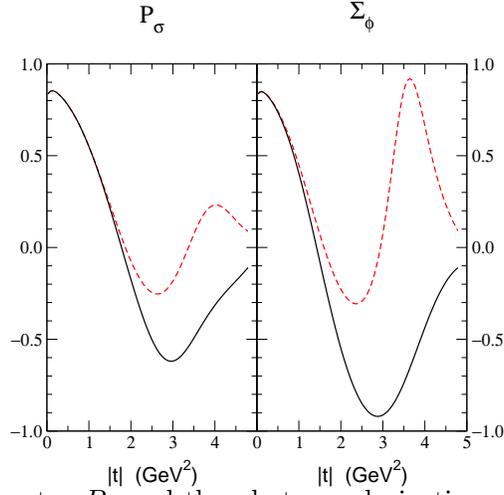}
\caption{The parity asymmetry $P_\sigma$ and the photon polarization
asymmetry $\Sigma_\phi$ in $\phi$ photoproduction at $E_\gamma = 3.6$ GeV
with the full amplitude. 
Notations are the same as in Fig. \protect\ref{fig:dm2_8}.}
\label{fig:asym3_6}
\end{figure}

This difference can be readily seen in the photon polarization asymmetry
$\Sigma_\phi$ and the parity asymmetry $P_\sigma$ which are defined as
\begin{eqnarray}
\Sigma_\phi &\equiv&
\frac{\sigma_\parallel - \sigma_\perp}{\sigma_\parallel +
\sigma_\perp} = \frac{1}{p_\gamma} \frac
{W^L(0,\frac{\pi}{2}, \frac{\pi}{2})- W^L(0,\frac{\pi}{2}, 0)}
{W^L(0,\frac{\pi}{2}, \frac{\pi}{2})+ W^L(0,\frac{\pi}{2}, 0)},
\nonumber \\
P_\sigma &\equiv& \frac{\sigma^N - \sigma^U}{\sigma^N + \sigma^U}
= 2 \rho_{1-1}^1 - \rho^1_{00},
\end{eqnarray}
where $W^L$ is the decay angular distribution of the $\phi$ meson,
$\sigma_\parallel$ ($\sigma_\perp$) is the cross sections for
symmetric decay particle pairs produced parallel (normal) to the
photon polarization plane and $\sigma^N$ ($\sigma^U$) is the
contribution of natural (unnatural) parity exchanges to the cross
section.
The details on the definitions and physical meanings of those
asymmetries can be found, e.g., in Refs. \cite{SSW70,BCGG72}.
Shown in Fig. \ref{fig:asym3_6} are our results for the asymmetries
at $E_\gamma = 3.6$ GeV.
As can be seen in Fig. \ref{fig:asym3_6}, measuring those asymmetries
can be used to constrain the $\phi NN$ coupling and can test the vector
meson dominance hypothesis.

\begin{figure}[t]
\centering
\epsfig{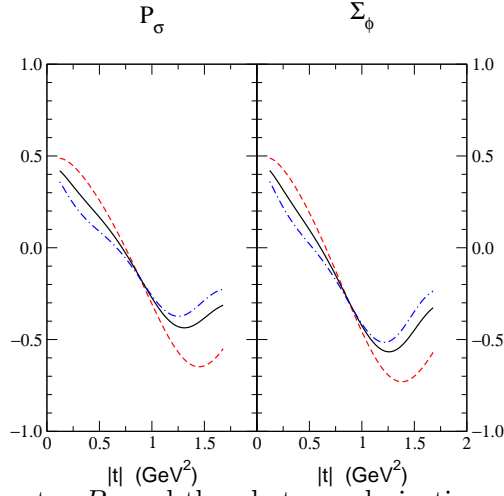}
\caption{The parity asymmetry $P_\sigma$ and the photon polarization
asymmetry $\Sigma_\phi$ in $\phi$ photoproduction at $E_\gamma = 2.0$ GeV
with the full amplitude with $g_{\phi NN}^{} = -3.0$. 
The dashed lines are without the knockout mechanisms ($B^2=0$).
The solid and dot-dashed lines are with $B^2=1$ \% and $2$ \%, respectively.}
\label{fig:asym2_0}
\end{figure}

As discussed in Refs. \cite{TOY97,TOYM98} and can be verified from
Fig.~\ref{fig:dsdt}, however, the direct knockout mechanisms could be
investigated only at lower energies, i.e., near the threshold.
Therefore, we compute the polarization asymmetries at $E_\gamma = 2.0$
GeV in Fig.~\ref{fig:asym2_0}.
The dashed lines are obtained without the knockout mechanisms, i.e.,
with $B^2 = 0$, while the solid and dot-dashed lines are with
$B^2 = 1$~\% and $2$~\%, respectively.
The results are dependent on the phases of $b_{0,1}$ and
the phases of $b_0$ and $b_1$ are assumed to be positive in
Fig.~\ref{fig:asym2_0} for simplicity.
We have used $g_{\phi NN}^{} = -3.0$ in the calculation.
The difference of the results depending on $B^2$ can be seen especially
at large $|t|$ region.
Although the difference is not as large as in Fig.~\ref{fig:asym3_6},
it can be used for to study the $uud$ knockout process
[Fig.~\ref{fig:diag}(f)] and might be tested experimentally.

In summary, we have investigated the vector meson density matrix and the
associated polarization asymmetries in $\phi$ photoproduction at low
energies.
As the models for $\phi$ photoproduction, we have considered the Pomeron
and pseudoscalar $(\pi,\eta)$ exchanges, nucleon pole terms, and
the direct knockouts.
We found that the recent CLAS data on $\phi$ photoproduction can be
explained by a rather large $\phi NN$ coupling, i.e. $g_{\phi NN} =
-3.0$, which supports the result of Ref. \cite{Lage00}.
This value is larger than the expectations based on the OZI rule, but is
consistent with the values estimated in electromagnetic nucleon form
factors and baryon-baryon scattering.
As another test for constraining $g_{\phi NN}$ we computed the vector
meson density matrix and the polarization asymmetries.
It was shown that they are strongly dependent on the $\phi NN$
coupling constant.
The results are different from the predictions of Ref. \cite{ZSA01}
which includes the nucleon resonances decaying into $\phi
N$ channel within an SU(3) quark model.
Thus the polarization asymmetries can be used to test the models for the
nucleon pole terms and the nucleon resonances.
Finally, we have shown that they can be used to probe the hidden
strangeness content of the nucleon.
Those observables could be measured at current experimental facilities
such as Thomas Jefferson National Accelerator Facility and LEPS of
SPring-8.
On the theoretical side, further investigations are needed for clarifying
the model dependence of our results and understanding the
contributions from the loop diagrams which give an effective $\phi NN$
coupling through the rescattering processes.

\acknowledgements

We are grateful to V. Burkert, M. Fujiwara, T.-S. H. Lee, and E. Smith
for fruitful discussions.
This work was supported in part by the Brain Korea 21 project of Korean
Ministry of Education and the International Collaboration Program of
KOSEF under Grant No. 20006-111-01-2.

\bigskip

\end{document}